# High third-order Kerr optical nonlinearity in BiOBr 2D films measured by the Z-scan method

Linnan Jia,[1] Dandan Cui,[2] Jiayang Wu,[1] Haifeng Feng,[3] Yunyi Yang,[4] Tieshan Yang,[4] Yang Qu,[1] Yi Du,[3] Weichang Hao,[2] Baohua Jia,[1,4, a] and David J. Moss[1, b]

[1]*Centre for Micro-Photonics, Swinburne University of Technology, Hawthorn, VIC 3122, Australia*
[2]*School of Physics, and BUAA-UOW Joint Research Centre, Beihang University, Beijing 100191, China*
[3]*Institute for Superconducting and Electronic Materials, and UOW-BUAA Joint Research Centre, University of Wollongong, Wollongong, NSW 2500, Australia*
[4]*Centre for Translational Atomaterials, Swinburne University of Technology, Hawthorn, VIC 3122, Australia*

\* E-mail: a) bjia@swin.edu.au ; b) dmoss@swin.edu.au;

We investigate the nonlinear optical properties of BiOBr nanoflakes – a novel two-dimensional (2D) layered material from the Bismuth oxyhalide family. We measure the nonlinear absorption and Kerr nonlinearity of BiOBr nanoflakes at both 800 nm and 1550 nm via the Z-Scan technique. We observe a large nonlinear absorption coefficient $\beta \sim 10^{-7}$ m/W as well as a large Kerr coefficient $n_2 \sim 10^{-14}$ m$^2$/W. We also observe strong dispersion in $n_2$, with it reversing sign from negative to positive as the wavelength varies from 800 nm to 1550 nm. In addition, we characterize the thickness-dependence of the nonlinear optical properties of BiOBr nanoflakes, observing that both the magnitudes of $\beta$ and $n_2$ increase for very thin flakes. Finally, we integrate BiOBr nanoflakes onto silicon integrated waveguides and characterize the linear optical properties of the resulting hybrid integrated devices, with the measurements agreeing with calculated parameters using independent ellipsometry measurements. These results verify the strong potential of BiOBr as an advanced nonlinear optical material for high-performance hybrid integrated photonic devices.

**I. INTRODUCTION**

All-optical signal processing based on nonlinear photonic devices has provided a competitive solution to realize ultrafast information processing in modern communications systems[1-4] with its broad operation bandwidth, low power consumption, and potentially reduced cost. As the key building blocks for implementing nonlinear photonic devices, advanced optical materials with superior nonlinear properties have been widely investigated.[1-15] Among them, two-dimensional (2D) layered materials such as graphene,[5,6] graphene oxide (GO),[7-9] transition metal dichalcogenides (TMDCs),[10-12] and black phosphorus (BP)[13,14] have attracted significant interest in recent years. Their remarkable optical properties, such as ultrahigh Kerr optical nonlinearities, strong nonlinear absorption, significant material anisotropy, and layer-dependent material properties have already enabled diverse new photonic devices that are fundamentally different from those based on traditional bulk materials.[15-17]

In addition to the 2D materials mentioned above, other layered materials have attracted interest.[18-21] Bismuth oxyhalides, i.e., BiOX (X = Cl, Br, I), which consist of [Bi$_2$O$_2$]$^{2+}$ slabs interleaved with double halogen atoms with weak van der Waals interactions among adjacent slabs of halogen, have been explored as a new group of advanced 2D layered materials.[22-24] Their unique open-layer crystal structure enables self-built internal static electric fields that lead to an effective separation of photoinduced charge carriers, which provides BiOX with superior photocatalytic behavior[24-26] and excellent nonlinear optical performance.[27,28] In Refs.,[27,28] the strong nonlinear optical properties of BiOCl, in terms of both nonlinear absorption and Kerr nonlinearity, have been demonstrated at 515 nm and 800 nm.

In this paper, we report on the nonlinear optical properties of BiOBr nanoflakes - an important member of the BiOX family. We measure the nonlinear absorption and Kerr nonlinearity of BiOBr nanoflakes at both 800 nm and 1550 nm via the Z-Scan technique.[29] Experimental results show that BiOBr nanoflakes have a strong nonlinear two-photon absorption coefficient $\beta$ of $\sim 10^{-7}$ m/W and a high Kerr nonlinearity coefficient $n_2 \sim 10^{-14}$ m$^2$/W at both wavelengths − both much higher than those of BiOCl.[27,28] We also note that the $n_2$ of BiOBr reverses sign from negative to positive as the wavelength is changed from 800 nm to 1550 nm. Moreover, unlike previous reports[27,28] where BiOX samples were either formed in dispersed nanosheets in solution or randomly distributed nanoflakes with nonuniform thickness, we prepare BiOBr nanoflakes with highly uniform thickness and characterize their nonlinear optical response as a function of thickness. It is found that the magnitude of $\beta$ and $n_2$ of the BiOBr nanoflakes increases at very thin flake thicknesses. We also integrate the BiOBr nanoflakes onto silicon integrated waveguides and measure the insertion loss of the hybrid integrated devices, with the extracted waveguide propagation loss showing good agreement with mode simulations based on ellipsometry measurements. These results confirm the strong potential of BiOBr as a promising nonlinear optical material for high-performance hybrid integrated photonic devices.

## II. MATERIAL PREPARATION AND CHARACTERIZATION

Fig. 1(a) illustrates the atomic structure of BiOBr crystals, where [Bi$_2$O$_2$]$^{2+}$ slabs are interleaved with double Br atoms to form a layered structure.[22,24] The self-built internal static electric field resulting from asymmetric charge distribution between the [Bi$_2$O$_2$]$^{2+}$ and Br layers leads to an effective separation of photoinduced electro-hole pairs,[24] which is very useful for the enhancement of the nonlinear optical response.[27,28]

The bulk BiOBr was synthesized by a typical hydrothermal method.[30,31] BiOBr nanoflakes with different thicknesses were mechanically exfoliated from the bulk crystals onto glass substrates using adhesive tape. The morphology images and thickness profiles of the prepared BiOBr nanoflakes were characterized by atomic force microscopy (AFM), as shown in Fig. 1(b). The measured thicknesses of the samples in (i) − (iv) were ~30 nm, ~75 nm, ~110 nm, and ~140 nm, respectively. Compared with dispersed nanosheets in solution or randomly distributed nanoflakes with a large thickness variation (>100 nm) in previous reports,[27,28] our prepared samples had much better thickness uniformity (variation < 5 nm). Since 2D layered materials usually have thickness-dependent optical properties,[32-34] precise control of their thickness is critical to minimize the variation of optical properties induced by thickness, allowing accurate characterization of the optical properties of BiOBr nanoflakes.

The linear absorption of BiOBr in the visible to infrared wavelength range was measured by ultraviolet-visible (UV-vis) spectrometry (Fig. 2(a)). There is a strong linear absorption edge near ~ 450 nm, which corresponds to a photon energy of ~ 2.76 eV, in agreement with the known bandgap of BiOBr of ~2.73 eV for bulk crystals and ~2.82 eV for monolayers.[26] As expected, the BiOBr nanoflakes exhibited very weak linear absorption at longer wavelengths - both at 800 nm (~1.55 eV) and 1550 nm (~0.80 eV). The transmittance spectra of BiOBr nanoflakes with different thicknesses is shown in Fig. 2(b). High transmittance (> 60%) for wavelengths from 1100 nm to 1600 nm is observed for all samples. In particular, the transmittance reached 90% for the 30-nm-thick BiOBr nanoflakes. We also characterized the BiOBr nanoflakes via Raman spectroscopic measurements (Fig. 2(c)) with a pump laser wavelength of ~532 nm. Two phonon modes of A$_{1g}$ (~113.2 cm$^{-1}$) and E$_g$ (~160.4 cm$^{-1}$) are observed for all samples, which is consistent with previous reports[35,36] and verifies the high quality of our BiOBr nanoflakes. We also note that the intensities of both the $A_{1g}$ and $E_g$ peaks increase with increasing sample thickness. The thickness-dependent variation of Raman signal intensity has also been observed for other layered materials.[11,37] Fig. 2(d) shows the in-plane (TE-polarized) refractive index (*n*) as well as extinction coefficient (*k*) of BiOBr measured by spectral ellipsometry,[38] for a sample thickness of ~ 1 μm. Since the out-of-plane (TM polarized) response of the thin samples is much weaker,[38,39] we could only measure the in-plane (TE polarized) *n* and *k* of the BiOBr thin flakes. A high refractive index of ~2.2 is obtained in the telecommunications band. A relatively high extinction coefficient *k* (~ 0.2) is also observed, which can be reduced by optimizing the preparation and properties of BiOBr, including crystal facet tailoring[40,41] as well as thickness tuning.[23,26]

## III. Z-SCAN MEASUREMENTS

The nonlinear absorption and refraction of the prepared BiOBr samples were characterized via Z-scan techniques,[29] with the experimental setup shown in Fig. 3. Femtosecond pulsed lasers, with center wavelengths at ~800 nm and ~1550 nm, were used to excite the samples. The 1550 nm femtosecond pulses were generated via an optical parametric oscillator (OPO) pumped by the Ti:sapphire femtosecond pulsed laser (Coherent, Chameleon) at 800 nm. The repetition rate and pulse duration were ~80 MHz and ~140 fs, respectively. A half-wave plate combined with a linear polarizer was employed as a power attenuator to control the power of incident light. A beam expansion system was comprised of a −25 mm concave lens and 150 mm convex lens to expand the light beam, which was then focused by an objective lens (10×, 0.25NA) to achieve a low beam waist with a focused spot size much smaller than the sample size, at ~1.6 μm and ~3.1 μm for the 800 nm and 1550 nm pulsed lasers, respectively. The prepared samples were oriented perpendicular to the beam axis and translated along the Z-axis with a highly precise linear motorized stage. A high definition charge-coupled-device (CCD) imaging system was used to align the light beam to the target sample. Two power detectors (PDs) were employed to collect the transmitted light powers.

We performed the Z-scan measurements in two stages − an open aperture (OA) measurement where all light transmitted through the sample was collected by the PD and a closed aperture (CA) measurement where a small aperture was placed before the PD so that only part of the on-axis transmitted beam was collected. For the OA measurements, the change of optical transmittance was caused by the nonlinear absorption. For CA measurement, the change of optical transmittance resulted from both nonlinear absorption and nonlinear phase shift induced by Kerr nonlinearity. Therefore, the ratio of the CA result to the OA result excludes the impact of nonlinear absorption and only reflects the nonlinear phase shift induced by the Kerr nonlinearity.

Figs. 4(a) and (b) show the Z-scan results of a ~140 nm thick BiOBr sample at 800 nm and 1550 nm, respectively, with (i) showing the OA results and (ii) the CA results. The irradiance laser intensities at 800 nm and 1550 nm were 0.202 GW/cm$^2$ and 0.442 GW/cm$^2$, respectively. Typical reverse saturation absorption (RSA)[11,13,27] behavior is observed in the OA curves at both wavelengths, indicating strong nonlinear absorption in the BiOBr sample. The slight deviation in experimental data from the standard symmetric OA curves is probably due to scattering from minor particles on the BiOBr samples[42,43] or the irregularities and/or asymmetry of the input laser beam profile.[44] Considering the relation between the photon energy of the excitation laser (~1.55 eV at 800 nm and ~0.8 eV at 1550 nm) and the bandgap of BiOBr (~2.73 eV for the bulk and ~2.82 eV for the monolayer),[26] the nonlinear absorption is associated with two-photon absorption (TPA) at 800 nm and multi-photon absorption (MPA) at 1550 nm. To extract the nonlinear coefficient $\beta$ of BiOBr, we fit the measured OA results in Figs. 4(a-i) and 4(b-i) by:[14,33]

$$T_{\text{OA}}(z) \simeq 1 - \frac{1}{2\sqrt{2}} \frac{\beta I_0 L_{\text{eff}}}{(1+x^2)}, \qquad (1)$$

where $T_{\text{OA}}(z)$ is the normalized optical transmittance of OA measurement, and $x = z/z_0$, with $z$ and $z_0$ denoting the sample position relative to the focus and the Rayleigh length of the laser beam, respectively; $L_{\text{eff}} = (1-e^{-\alpha_0 L})/\alpha_0$ is the effective sample thickness, with $\alpha_0$ and $L$ denoting the linear absorption coefficient and the sample thickness, respectively and $I_0$ is the irradiance intensity at focus. The fit $\beta$ values at 800 nm (Fig. 4(a-i)) and 1550 nm (Fig. 4(b-i)) are $\sim$1.869×10$^{-7}$ m/W and $\sim$1.554 × 10$^{-7}$ m/W, respectively. The fit $\beta$ at 1550 nm is slightly lower than that at 800 nm, probably due to the lower efficiency for MPA (at 1550 nm) as compared with TPA (at 800 nm).

The Kerr nonlinearity of the BiOBr sample at both 800 nm and 1550 nm was characterized by the CA measurements, shown in Figs. 4(a-ii) and 4(b-ii) at 800 nm and 1550 nm, respectively. Note that although we refer to them as CA, they actually show the ratio of the CA result to the OA result (automatically processed by software) that reflects the nonlinear phase shift induced by the Kerr nonlinearity. In Fig. 4(a-ii), prominent peak-valley transmittance is observed for the CA measurement at 800 nm, which reflects optical self-defocusing in BiOBr nanoflakes and corresponds to a negative Kerr coefficient $n_2$. In contrast, valley-peak transmittance is observed at 1550 nm in Fig. 4(b-ii), which is a reflection of optical self-focusing, resulting in a positive $n_2$. To extract the $n_2$ of BiOBr, the measured CA results in Figs. 4(a-ii) and 4(b-ii) are fit by:[29]

$$T_{\text{CA}}(z, \Delta\Phi_0) \simeq 1 + \frac{4\Delta\Phi_0 x}{(x^2+9)(x^2+1)}, \qquad (2)$$

where $T_{\text{CA}}(z, \Delta\Phi_0)$ is the normalized optical transmittance of CA measurement. $\Delta\Phi_0 = 2\pi n_2 I_0 L_{\text{eff}}/\lambda$ is the nonlinear phase shift, with $\lambda$ denoting the laser center wavelength. The fit Kerr coefficients ($n_2$) values at 800 nm (Fig. 4(a-ii)) and 1550 nm (Fig. 4(b-ii)) are $\sim$-1.737 ×10$^{-14}$ m$^2$/W and $\sim$3.824 ×10$^{-14}$ m$^2$/W, respectively. As expected, the values of $n_2$ at 800 nm and 1550 nm have opposite signs. The transition from negative $n_2$ at 800 nm to positive $n_2$ at 1550 nm can be attributed to the dispersion of $n_2$ associated with the two-photon bandgap where $n_2$ is positive when the excitation photon energy (at 1550 nm) is below the TPA band edge (half-bandgap) while it may become negative when the photon energy (at 800 nm) is between the one-photo absorption and TPA edges.[45,46] The fit values of $n_2$ (at 800 nm and 1550 nm) are almost three orders of magnitude larger than single crystalline silicon,[1] which highlights the strong Kerr nonlinearity of the BiOBr nanoflakes. In Table 1, we compare the measured $\beta$ and $n_2$ of BiOBr with other 2D layered materials, observing that BiOBr nanoflakes exhibit a large $\beta$ on the order of 10$^{-7}$ m/W, which is much higher than many other layered materials, and in particular is two orders of magnitude higher than BiOCl. The Kerr coefficient $n_2$ is on the order of 10$^{-14}$ m$^2$/W, which is close to that of graphene and GO, and is more than one order of magnitude higher than TMDCs and BiOCl. These results confirm the superior nonlinear optical properties of BiOBr nanoflakes as an advanced nonlinear optical material.

In Table 1, we also calculate the nonlinear figure of merit (FOM) given by:

$$\text{FOM} = n_2 / (\beta \times \lambda), \qquad (3)$$

where $n_2$, $\beta$, and $\lambda$ are the Kerr coefficient, the nonlinear absorption coefficient, and the light wavelength, respectively. We note that the concept of the nonlinear FOM was originally proposed purely in the context of a limitation for nonlinear materials having a positive $n_2$.[47] However, 2D layered films have demonstrated very complex behavior, including both a negative $n_2$ and negative nonlinear absorption (e.g., due to saturable absorption), and it is not clear that the conventional FOM concept is relevant or useful in these cases.

Table 1. Comparison of $\beta$, $n_2$, and figure of merit (FOM) of various 2D layered materials

| Material | Laser parameter | Thickness (nm) | $\beta$ (m/W) | $n_2$ (m$^2$/W) | FOM | Reference |
|---|---|---|---|---|---|---|
| Graphene | 1550 nm, 100 fs | 5-7 layers | $9 \times 10^{-8}$ | $-8 \times 10^{-14}$ | -0.574 | 6 |
| GO | 800 nm, 100 fs | $2 \times 10^3$ | $4 \times 10^{-7}$ | $1.25 \times 10^{-13}$ | 0.391 | 7 |
| GO | 1560 nm, 67 fs | $1 \times 10^3$ | $-5.25 \times 10^{-8}$ | $4.5 \times 10^{-14}$ | -0.5 | 9 |
| MoS$_2$ | 1064 nm, 25 ps | $2.5 \times 10^4$ | $-3.8 \times 10^{-11}$ | $1.88 \times 10^{-16}$ | -4.649 | 10 |
| MoS$_2$ | 1030 nm, 350 fs | 50 | $4.99 \times 10^{-9}$ | N/A | N/A | 48 |
| WS$_2$ | 1040 nm, 340 fs | 57.9 | $1.81 \times 10^{-8}$ | $-3.36 \times 10^{-16}$ | -0.018 | 11 |
| WSe$_2$ | 1040 nm, 340 fs | 25.1 | $2.14 \times 10^{-8}$ | $-1.71 \times 10^{-15}$ | -0.077 | 11 |
| BP | 800 nm, 100 fs | 30-60 | $4.5 \times 10^{-10}$ | $6.8 \times 10^{-13}$ | $1.889 \times 10^3$ | 13 |
| BP | 1030 nm, 140 fs | 15 | $5.845 \times 10^{-6}$ | $-1.635 \times 10^{-12}$ | -0.272 | 14 |
| BiOCl | 800 nm, 100 fs | 20-140 | $4.25 \times 10^{-9}$ | $3.8 \times 10^{-15}$ | 1.118 | 28 |
| BiOBr | 800 nm, 140 fs | 30 | $6.011 \times 10^{-7}$ | $-3.155 \times 10^{-14}$ | -0.066 | This work |
| BiOBr | 800 nm, 140 fs | 140 | $1.869 \times 10^{-7}$ | $-1.737 \times 10^{-14}$ | -0.116 | This work |
| BiOBr | 1550 nm, 140 fs | 140 | $1.554 \times 10^{-7}$ | $3.824 \times 10^{-14}$ | 0.159 | This work |

Fig. 5 shows the OA and CA results for BiOBr samples with different thicknesses. We used the femtosecond pulsed lasers centered at ~800 nm to excite the samples. The irradiance laser intensity was 0.202 GW/cm$^2$. Table 2 summarizes the fit $\beta$ and $n_2$ obtained from Fig. 5. Fig. 6(a) depicts the fit $\beta$ for different thicknesses. It can be seen that $\beta$ for the BiOBr nanoflakes is highly thickness dependent, where it increases from ~1.869×10$^{-7}$ m/W to ~6.011×10$^{-7}$ m/W when the sample thickness decreases 140 nm to 30 nm. The dependence of $\beta$ on sample thickness is likely induced by localized defects in BiOBr nanoflakes, which would lead to more scattering and energy loss for thicker BiOBr nanoflakes, thus resulting in a decreased $\beta$.[33,34] It is also worth mentioning that a large $\beta$ as high as ~ 6.011 × 10$^{-7}$ m/W is obtained for the 30-nm-thick BiOBr sample, which is more than an order of magnitude higher than other typical 2D materials such as graphene[6] and TMDCs.[10,11] Such a strong nonlinear absorption is highly desirable for the implementation of high-performance optical limiters to realize pulse shaping, mode locking, and sensor focal-arrays.[5,49-51] The extracted $n_2$ for different thicknesses is depicted in Fig. 6(b), where a strong thickness-dependent $n_2$ is observed, with the highest $n_2$ value of ~3.155×10$^{-14}$ m$^2$/W being obtained for the 30-nm-thick BiOBr nanoflake. The value of $n_2$ decreases to ~1.737 × 10$^{-14}$ m$^2$/W for a thickness of 140 nm. Note that the fit $n_2$ of BiOBr nanoflakes is close to the reported values of graphene,[6] which is much higher than common materials for integrated platforms, such as silicon and silicon nitride.[1]

Table 2. Nonlinear parameters of BiOBr nanoflakes with different thicknesses

| Thickness (nm) | Laser parameter | $\beta$ (m/W) | $n_2$ (m$^2$/W) | FOM |
|---|---|---|---|---|
| 30 | 800 nm, 140 fs | $(6.011 \pm 0.181) \times 10^{-7}$ | $(-3.155 \pm 0.192) \times 10^{-14}$ | -0.066 |
| 75 | 800 nm, 140 fs | $(5.548 \pm 0.251) \times 10^{-7}$ | $(-2.732 \pm 0.345) \times 10^{-14}$ | -0.062 |
| 110 | 800 nm, 140 fs | $(2.593 \pm 0.606) \times 10^{-7}$ | $(-1.771 \pm 0.058) \times 10^{-14}$ | -0.086 |
| 140 | 800 nm, 140 fs | $(1.869 \pm 0.687) \times 10^{-7}$ | $(-1.737 \pm 0.035) \times 10^{-14}$ | -0.116 |
| 140 | 1550 nm, 140 fs | $(1.554 \pm 0.053) \times 10^{-7}$ | $(3.824 \pm 0.032) \times 10^{-14}$ | 0.159 |

## IV. INTEGRATION ON SILICON PHOTONIC DEVICES

In this section, we characterize the BiOBr nanoflakes integrated in 220-nm-thick silicon-on-insulator (SOI) waveguides[52,53] on a 2-μm-thick buried oxide (BOX) layer. Photolithography using stepper 248 nm deep ultraviolet patterning defined the device layout, with an inductively coupled plasma (ICP) etching process being used to etch the top silicon layer. Inverse taper couplers were employed to couple light into and out of the devices with lensed fibers. A 1.5-μm-thick silica layer was deposited by plasma enhanced chemical vapor deposition as an upper cladding layer. Finally, a window was opened down to the BOX layer via photolithography and reactive ion etching to allow the introduction of BiOBr nanoflakes.

BiOBr nanoflakes were transferred onto the silicon integrated waveguides using an all-dry transfer method.[33,54] We selected a thin film of polydimethylsiloxane (PDMS) as the transfer stamp which was adhered to a glass slide to facilitate its handling. BiOBr nanoflakes were first transferred to the PDMS stamp by mechanical exfoliation of bulk crystal using adhesive tape. The stamp was then attached to a three-axis manipulator next to the silicon waveguides. To identify and select the BiOBr nanoflakes, an optical microscope was employed to inspect the stamp surface. Once the desired nanoflake was selected, the manipulator was used to align it to the target device assisted by the optical microscope. To achieve good attachment between BiOBr nanoflake and the waveguide, the stamp was pressed against the waveguide top surface gently and was peeled off very slowly. The BiOBr nanoflake was then detached from the stamp and adhered to the silicon waveguides due to van der Waals forces.

The microscope image of a silicon integrated waveguide incorporated with BiOBr nanoflake is shown in Fig. 7(a). The width of the waveguide was ~500 nm. It can be seen that the BiOBr nanoflake is attached to the silicon integrated waveguide, with an overlap length of ~13 μm. The thickness of the BiOBr nanoflake is ~110 nm. The measured TE polarized insertion losses of the silicon waveguides with and without a 110-nm-thick BiOBr nanoflake are ~6.9 dB and ~18.6 dB, respectively. The insertion loss measurements were performed at a wavelength of 1550nm and a continuous-wave (CW) power level of 0 dBm. The butt coupling loss was ~ 3.0 dB each, or ~6.0 dB for both. According to our previously fabricated devices, the waveguide propagation loss for single-mode silicon nanowire waveguides with a cross section of 500 nm × 220 nm is about ~3 dB/cm,[55,56] which is much lower than that of the hybrid waveguide. Therefore, the propagation loss of the hybrid waveguide can be given by:

$$PL = (IL_{hybrid} - IL_{silicon}) / L_{hybrid}, \qquad (4)$$

where $IL_{hybrid}$ and $IL_{silicon}$ are the insertion losses of the hybrid waveguide and the silicon waveguide without BiOBr, respectively. $L_{hybrid}$ is the length of the BiOBr nanoflake on the silicon waveguide. Fig. 7(b) shows the TE and TM polarized waveguide propagation loss extracted from the measured insertion losses. We measured the hybrid waveguides with four different BiOBr thicknesses. The data points depict the average values obtained from the experimental results of three samples and the error bars illustrate the variations for different samples. It can be seen that the propagation loss of the hybrid waveguides increases with increasing BiOBr flake thickness. The lowest propagation loss of ~0.13 dB/μm was found to be for the 30-nm-thick BiOBr, which is larger than the reported values of graphene-Si[57, 58] and $MoS_2$-Si waveguides[59]. Moreover, the propagation loss for TM polarization is much higher than for TE polarization. Such a difference is mainly caused by mode overlap and can be used for implementing waveguide polarizers.[60, 61] We also perform mode analysis for the hybrid integrated waveguides using Lumerical FDTD commercial mode solving software. We used the in-plane (TE-polarized) $n$ and $k$ of 1-μm-thick BiOBr obtained from the ellipsometry measurements (the values in Fig. 2(d) at 1550 nm) in the FDTD simulation. Fig. 7(c) shows the TE mode profiles of the hybrid integrated waveguides with different BiOBr thicknesses. For comparison, the simulated waveguide propagation losses are shown in Fig. 7(b). It can be seen that the experimental propagation losses are close to the simulated losses, which reflects the stability of the prepared BiOBr nanoflakes. We also note that the simulated propagation losses based on the $n$ and $k$ of 1-μm-thick BiOBr are slightly higher than the experimental propagation losses, with the difference between them increasing as the BiOBr thickness decreases. This indicates that the intrinsic material loss actually increases with flake thickness, which could be attributed to a number of effects such as increased scattering loss and absorption induced by imperfect contact between the multiple layers as well as interactions between them.

## V. CONCLUSION

In summary, we investigate the nonlinear optical properties of BiOBr nanoflakes. We characterize the nonlinear absorption and Kerr nonlinearity of BiOBr nanoflakes at both 800 nm and 1550 nm via Z-Scan techniques, and obtain a large nonlinear absorption coefficient $\beta$ on the order of $10^{-7}$ m/W as well as a large Kerr coefficient $n_2$ on the order of $10^{-14}$ m$^2$/W. Dispersion in $n_2$ is observed, with $n_2$ changing sign from negative to positive when the wavelength is varied from 800 nm to 1550 nm. We also characterize the thickness-dependent nonlinear optical properties of BiOBr nanoflakes, finding that the magnitudes of $\beta$ and $n_2$ increase with decreasing thickness of the BiOBr nanoflakes. Finally, we integrate BiOBr nanoflakes into silicon integrated waveguides and measure their insertion loss, with the extracted waveguide propagation loss showing good agreement with mode simulations based on ellipsometry measurements. These results verify BiOBr as a promising nonlinear optical material for high-performance hybrid integrated photonic devices.

## ACKNOWLEDGEMENT

This work was supported by the Australian Research Council Discovery Projects Program (No. DP150102972 and DP190103186) and Beijing Natural Science Foundation (Z180007). We acknowledge Swinburne Nano Lab and Micro Nano Research Facility (MNRF) of RMIT University for the support in material characterization as well as Advanced Micro Foundry (AMF) Pte Ltd for the support in silicon device fabrication. We thank Dr. Tania Moein and Dr. Xiaorui Zheng for technical support, Dr. Chenglong Xu for assisting in optical characterization.

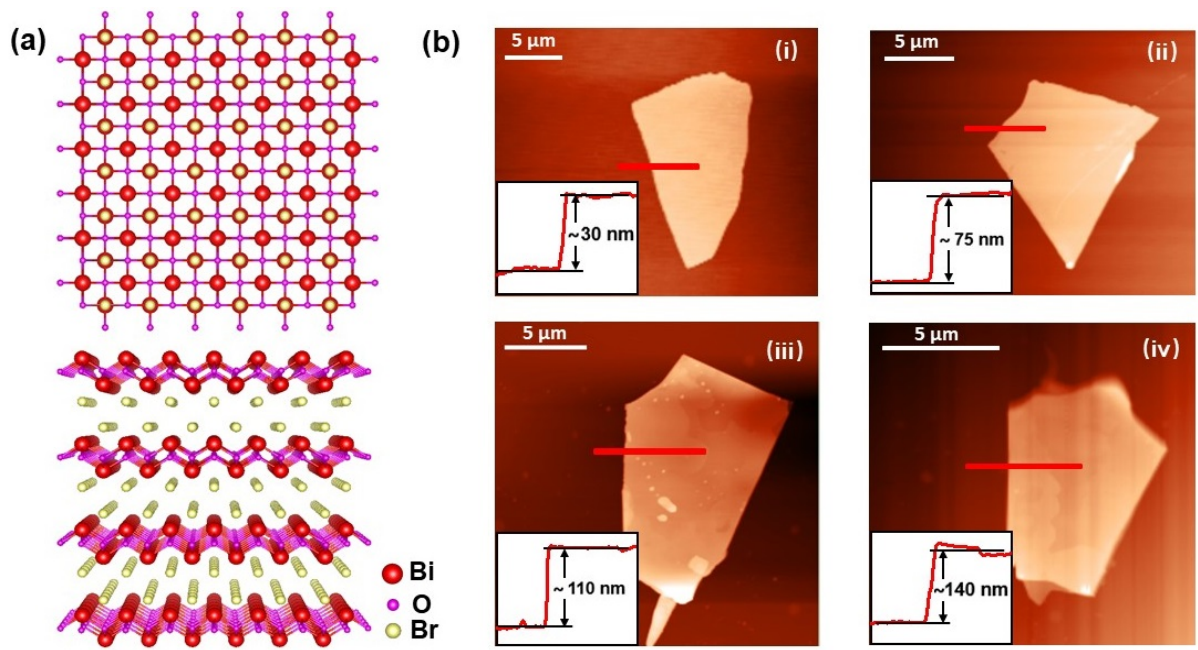

Fig. 1. (a) Schematic atomic structure of BiOBr. (b) AFM images and height profiles (insets) of exfoliated BiOBr nanoflakes with various thicknesses: (i) ~30 nm, (ii) ~75 nm, (iii) ~110 nm, (iv) ~140 nm.

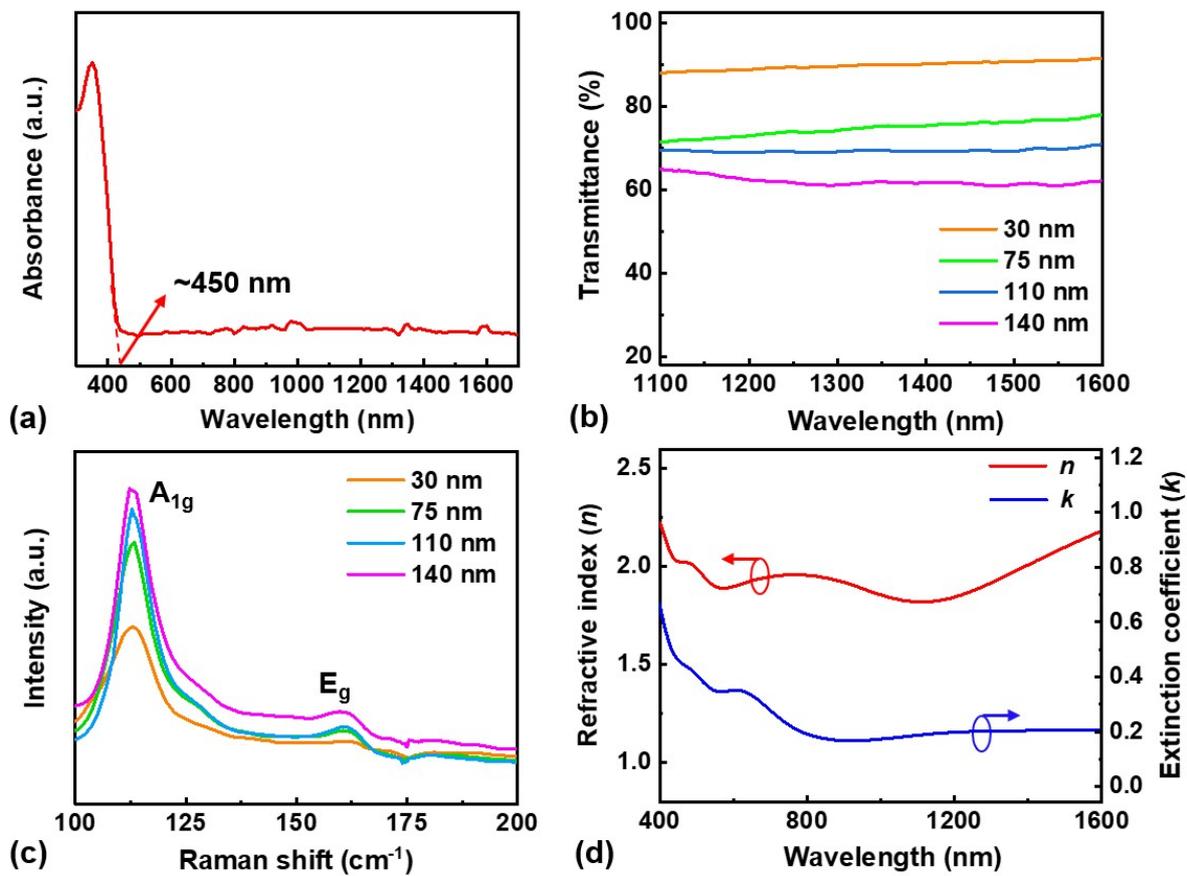

Fig. 2. (a) UV-vis absorption spectrum of BiOBr. (b) Measured linear transmittance spectra of BiOBr nanoflakes with different thicknesses. (c) Raman spectra of BiOBr nanoflakes with different thicknesses. (d) Measured refractive index ($n$) and extinction coefficient ($k$) of BiOBr.

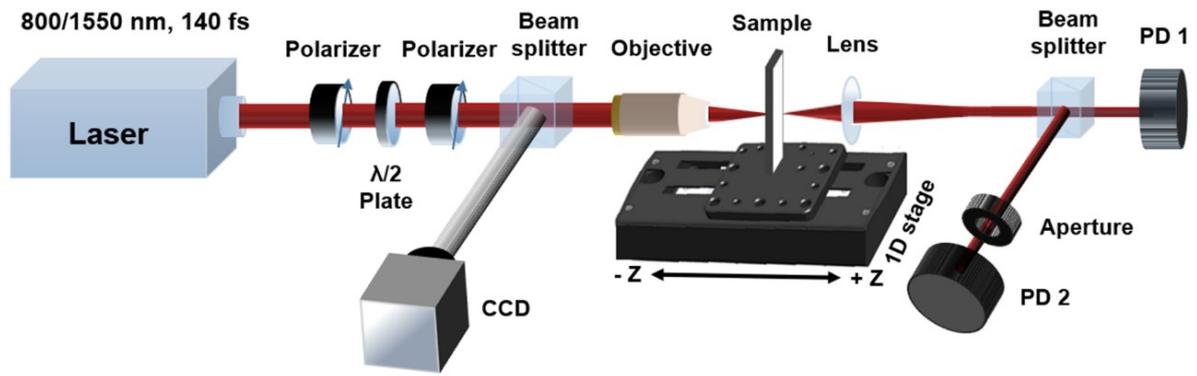

Fig. 3. Schematic illustration of Z-scan experimental setup. PD: power detector.

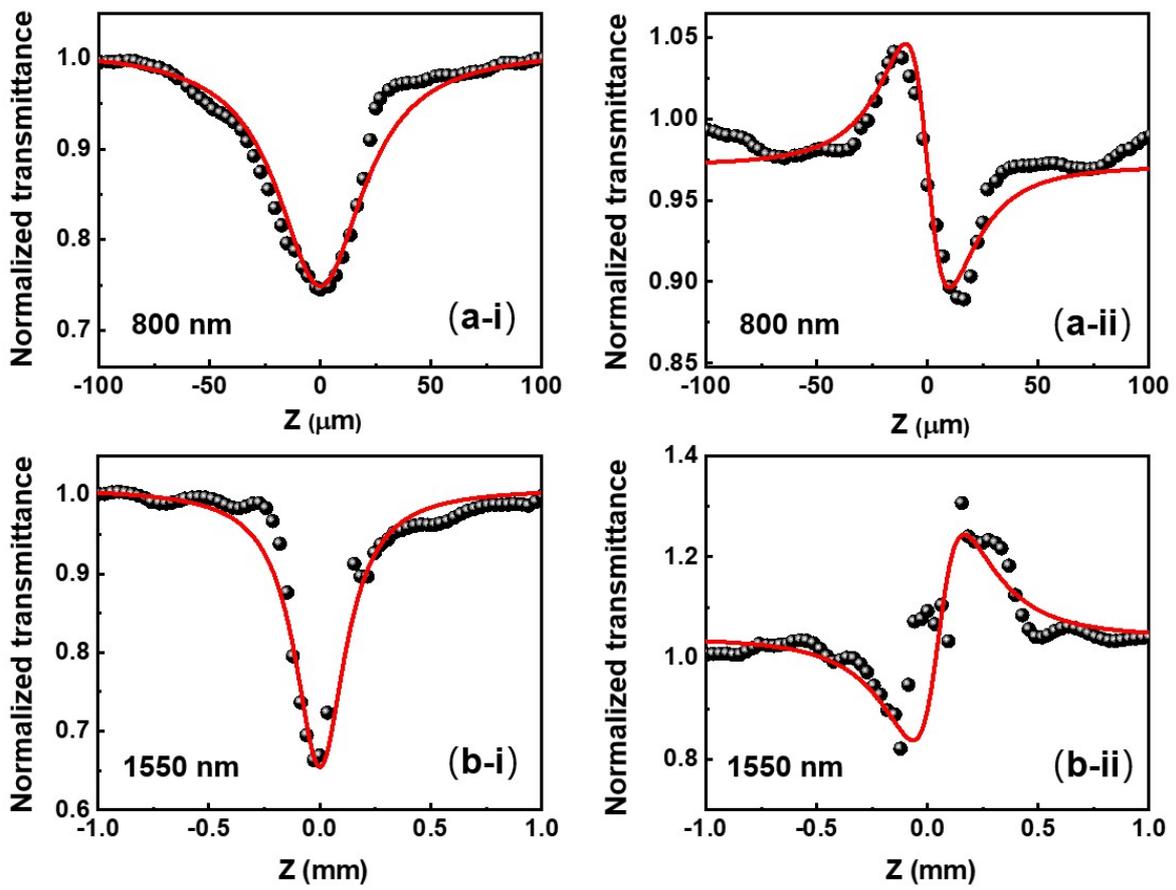

Fig. 4. (a) OA (i) and CA (ii) results at 800 nm. (b) OA (i) and CA (ii) results at 1550 nm. The thickness of the BiOBr sample is ∼140 nm. The measured and fit results are shown by black data points and red solid curves, respectively.

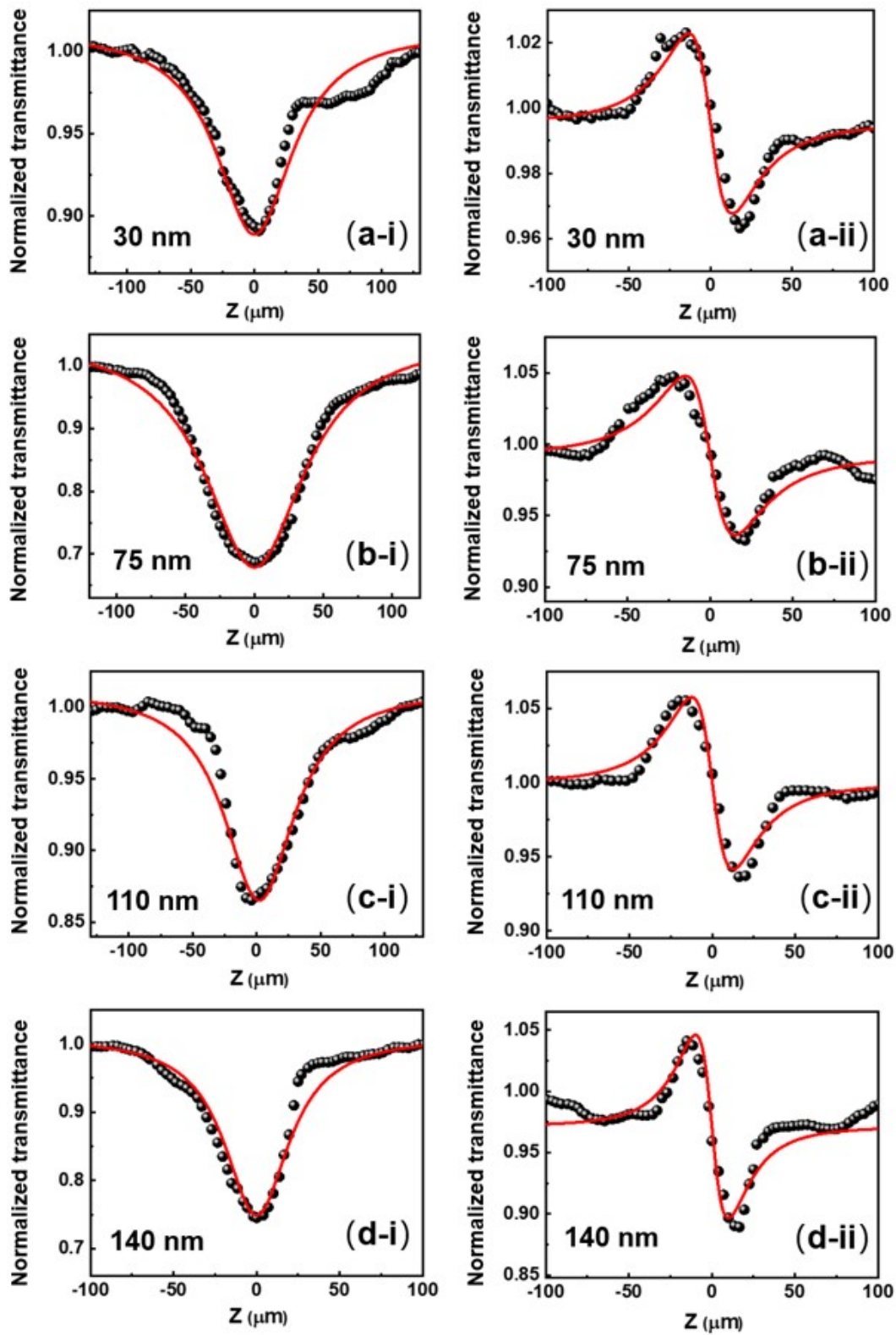

Fig. 5. Measured (black data points) and fit (red solid curves) Z-scan results for BiOBr samples with different thicknesses. The thicknesses of the BiOBr samples in (a) – (d) are ~30 nm, ~75 nm, ~110 nm, ~140 nm, respectively. In (a) – (d), (i) shows the OA results and (ii) shows the CA results.

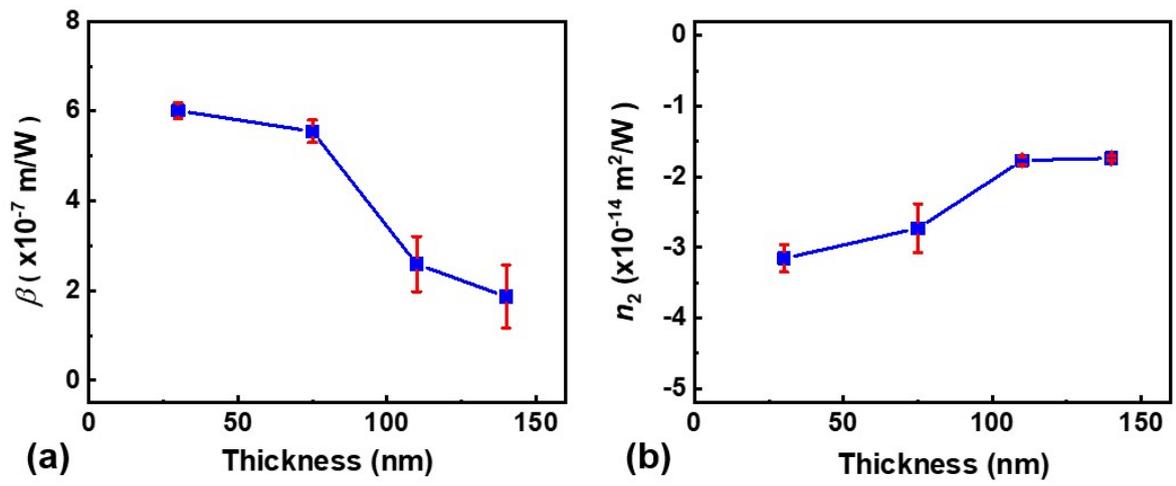

Fig. 6. Fit (a) nonlinear absorption coefficient $\beta$ and (b) Kerr coefficient $n_2$ for BiOBr samples with different thicknesses measured at λ=800nm.

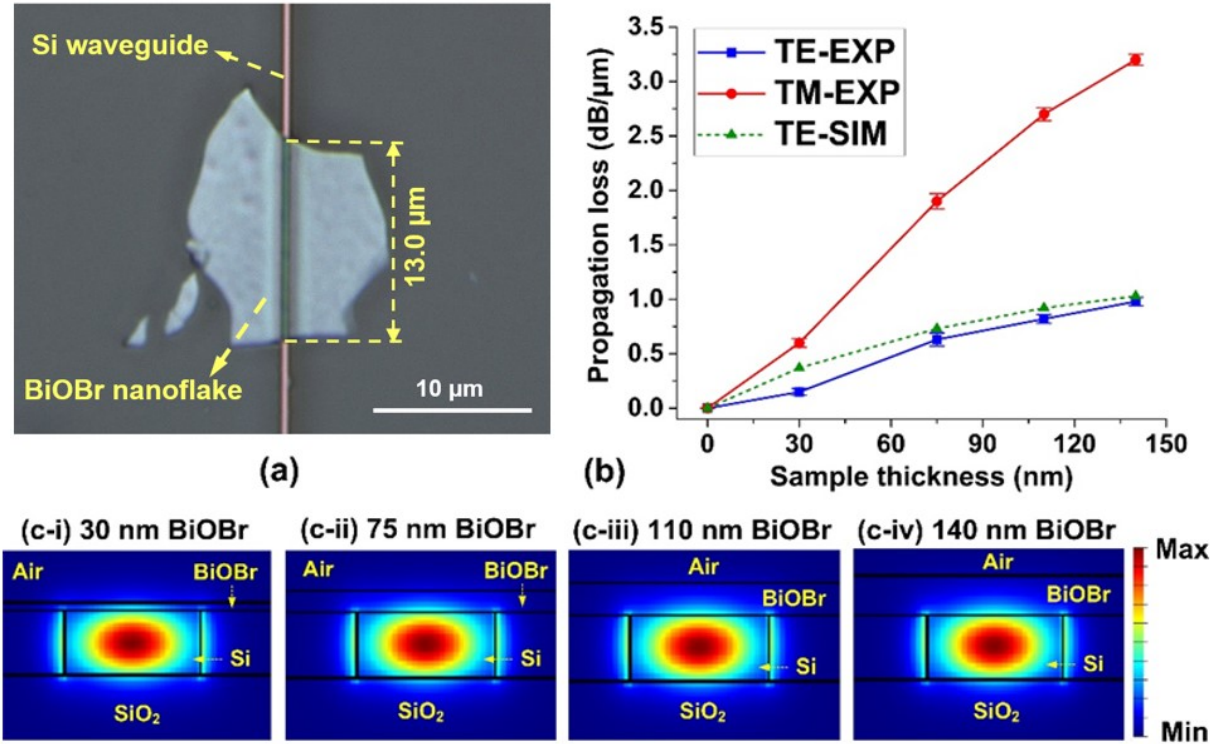

Fig. 7 (a) Microscope image of a silicon integrated waveguides incorporated with BiOBr nanoflake. (b) Measured and simulated waveguide propagation losses of the hybrid waveguides for different BiOBr thicknesses. (c) TE ($E_x$) mode profile of the hybrid integrated waveguide for different BiOBr thicknesses (i) 30 nm. (ii) 75 nm. (iii) 110 nm. (iv) 140 nm.